# Optical emission near a high-impedance mirror

Majid Esfandyarpour[1], Alberto G. Curto[1], Pieter G. Kik[1,2], Nader Engheta[3],

Mark L. Brongersma*[1]

1. Geballe Laboratory for Advanced Materials, Stanford University, 476 Lomita Mall, Stanford, California 94305
2. CREOL, The College of Optics and Photonics, University of Central Florida, Florida 32816, USA
3. Department of Electrical and Systems Engineering, University of Pennsylvania, Philadelphia, PA 19104

* To whom correspondence should be addressed. E-mail: Brongersma@stanford.edu.

**Abstract:**

Solid state light emitters rely on metallic contacts with high sheet-conductivity for effective charge injection. Unfortunately, such contacts also support surface plasmon polariton (SPP) excitations that dissipate optical energy into the metal and limit the external quantum efficiency. Here, inspired by the concept of radio-frequency (RF) high-impedance surfaces and their use in conformal antennas we illustrate how electrodes can be nanopatterned to simultaneously provide a high DC electrical conductivity and high-impedance at optical frequencies. Such electrodes do not support SPPs across the visible spectrum and greatly suppress dissipative losses while facilitating a desirable Lambertian emission profile. We verify this concept by studying the emission enhancement and photoluminescence lifetime for a dye emitter layer deposited on the electrodes.

**Description of Research:**

The development of high-efficiency and high-power solid-state light emitters has led to their dramatically increased use in lighting and display applications. Yet there are still significant opportunities to further boost performance. In particular, organic light emitting diodes (OLEDs) exhibit a low external quantum efficiency (EQE) which can largely be attributed to the excitation of dissipative SPPs in the metallic electrode[1,2]. These electrodes perform critical functions by facilitating carrier injection and serving as a mirror that redirects photons to the emitting surface. As such, they cannot be omitted from these devices and the SPP loss channel seems unavoidable. In this work, we provide a new solution for this longstanding fundamental challenge and break the usual dichotomy between high electrical and high optical performance of metallic electrodes.

Currently, optical energy dissipation in metallic electrodes is minimized by inserting a dielectric spacer layer between the metal and emitter layers. However, the use of such spacers increases the device's contact resistance and the probability of trapping light inside the high-index light emitting materials via total internal reflection[3]. Significant progress has been made in light extraction from LEDs [4–8]. Alternatively, researchers have attempted to reduce the SPP mode loss contribution by using wavelength-scale periodic gratings in the electrode that decouple excited SPPs into free-space radiation[9–12]. Although these type of structures can out-couple SPPs very efficiently, they give rise to a highly-directional and wavelength-dependent emission,[12] which is undesirable in many display and lighting applications. In light of these drawbacks, it would be highly desirable to tackle the dissipative loss in the metal at its root and remove the coupling of emitters to SPPs on the metallic contact without the need for spacer layers and complex light-extraction schemes.

Here, we illustrate how the detrimental excitation of SPPs can be avoided by implementing a high-impedance nanopatterned metal as the electrode. Our approach is inspired by works originally performed in radio frequency engineering to suppress the coupling of radiation to bound surface waves supported by a metallic ground plane[13]. The key is to create a high-impedance surface that does not support bound surface modes.

This is possible by patterning a metal film in such a way that the ratio of the electric over the magnetic field at the surface becomes extremely high. Such a surface was found to naturally avoid the undesirable "shorting" of an electric dipole antenna placed in close proximity to a high-conductivity metal film. It was also found to suppress the coupling of antenna emission to non-radiative bound surface modes. The history of high-impedance surfaces has its roots in the notion of electromagnetic "hard" and "soft" surfaces – a concept borrowed from acoustics and brought into the electromagnetic regime at microwave frequencies for RF antenna feed engineering[14–16] and for the design of artificial magnetic conducting surfaces suitable for conformal antennas[13,17]. Such surfaces were later proposed as a ground plane for thin absorbers[18,19]. Here, we transplant this concept into the visible domain with the specific goal of enhancing emission fromf optical emitters situated on such nano-patterned metasurfaces, effectively constructing optical "conformal" nanoemitters. With the recent appearance of several metasurface mirror designs operating as high-impedance surfaces in the visible spectral range,[20–25] it is logical to explore their application to enhance the emission from quantum dipole emitters located near a metallic electrode/reflector.

We start by showing how the radiative decay of a quantum emitter above a metal film can be modified through the introduction of a subwavelength corrugation. Figure 1a shows the electromagnetic fields excited by a dipole emitter placed in close proximity to a smooth metal surface serving as an electric mirror (EM), oscillating at a frequency corresponding to a free space wavelength of $\lambda_{em}$ = 600 nm. The emitter is seen to excite an SPP on the smooth metal surface, which subsequently decays in strength as it dissipates energy in the metal upon propagation. After introducing a subwavelength groove-array into the metal to form a high-impedance metasurface (HIM), there is no noticeable SPP excitation on the electrode surface (Fig. 1b). A similar observation can be made for an electric dipole that is oriented normal to the metal surface (Supplementary Fig. S1). This directly results in a greatly suppressed dissipative coupling into the metal and an increased far-field radiation (Supplementary Fig. S2). It can be seen that the far-field radiation is achieved without significantly affecting the broad angular radiation patterns of electric dipole emitters, which is desirable for many lighting and display applications (see Figs. 1c and S3a,b).

The absence of guided SPP modes on a properly designed HIM can be verified by launching a SPP along a smooth metal film in the direction of a finite-sized HIM region. In Fig. 1d, an SPP wave is traveling toward a set of ten-subwavelength grooves is seen to effectively decouple and radiate into the far-field in a mere 1-µm-long corrugated region. Notably, little backreflection of the SPP wave occurs. Simulations show that the SPP decoupling occurs efficiently across a wide frequency range while generating little SPP reflection or transmission for groove depths ranging from 50 nm to 120 nm (Supplementary Fig. S4).

The observations in Fig. 1 show that the HIM effectively transforms guided SPPs on a smooth metal surface into leaky waves. These leaky waves feature a lateral wave vector that is smaller than the free space wave vector and this causes them to rapidly decouple from the surface. This point can be substantiated by considering the band structure of the HIM (Fig. 2a) compared to the air light line and the SPP dispersion for a smooth, lossy silver surface. The patterned surface exhibits a region with no allowed SPP modes in the broad frequency range from 370 THz - 750 THz (400 nm - 810 nm) due to the high impedance of the structure. The guided modes at the upper and lower edge of this region correspond to plasmonic modes that feature a high magnitude of the magnetic field at the top of the corrugations and inside the grooves. The magnetic field profile $|H_z|$ for these two modes is shown in Fig. 2b and Fig. 2c. At frequencies between these extremes, no guided SPP modes are allowed, and consequently SPPs incident on a HIM are either reflected or radiated into free space, with the radiation channel dominating as a result of the high impedance of the surface as shown in Fig. S4. It is worth emphasizing that for the proposed subwavelength groove arrays (groove spacing $\lambda_{em}/4$) the formation of a bandgap is linked to the effective decoupling of quasi-guided waves from the surface. This behavior contrasts the behavior of typical plasmonic bandgap structures that feature shallow grooves with a spacing of half a SPP wavelength for which the formation of a bandgap is linked to SPP reflection in the plane of the metal surface, rather than the enhanced radiation away from the surface achieved here.

Next, we experimentally measure the transmission of SPPs across an array of ten subwavelength grooves. The targeted groove dimensions are 100 nm deep, 75 nm wide and the periodicity is 150 nm. To this end, an optically thick, silver film is deposited on a smooth glass substrate. Focused ion beam (FIB) milling is then used to carve a groove-array as shown in Fig. 2d. The groove-array is flanked on one side by a slit through the metal used to launch SPPs towards the groove-array. A decoupling groove is milled on the opposing side of the groove array to monitor the SPP transmission. Upon illumination of the excitation slit with a collimated white light source, both the groove array and the decoupling groove light up. The decoupling groove features a darker section directly behind the groove-array, indicating that as few as 10 grooves can effectively prevent the propagation of SPP waves by converting them to far-field radiation.

The SPP transmission spectrum is shown in Fig. 2e. The transmission coefficient is less than 0.2 for wavelengths in the range from 500 nm - 800 nm and increases abruptly at the edges of this spectral window. The onsets of high transmission occur close to the calculated upper and lower edges of the bandgap of the patterned surface. The low transmission in the bandgap region is due to the decoupling of SPPs by the HIM across the entire visible range, rather than due to Bragg backreflection of SPPs across a narrow bandwidth, as is observed for plasmonic bandgap structures based on shallow metallic gratings [26–28].

We can now exploit these HIMs to increase the external quantum efficiency of a light-emitting material. We cover a high-impedance surface region as shown in Fig. 3a with a 15-nm-thick layer of Rhodamine 6G (R6G) in polymethyl methacrylate (PMMA). The broad emission spectrum of R6G is peaked at 550 nm, inside the predicted band gap of the high impedance surface in Fig. 2a. In order to maintain a fixed density of emitters per unit area for different groove parameters, we first fill the grooves with $SiO_2$ to create a planar overcoat before spin-casting the dye-doped PMMA layer (see Methods). The emitter layer is then excited with a 485 nm pulsed laser. In order to minimize the effect of the excitation enhancement on fluorescence intensity, the laser is linearly polarized along the grooves (transverse electric or TE polarization) to avoid direct excitation of gap SPPs in the grooves that could result in large absorption enhancements in the dye layer. Photoluminescence (PL) maps are obtained by scanning the laser excitation spot across the sample while collecting the PL emission at each point with a confocal microscope. The maps are created while collecting the PL signal over the entire emission spectrum of the dyes.

As seen in emission maps for polarizations both parallel and perpendicular to the grooves (Figs. 3b and 3c), the emission intensity is relatively uniform across the patterned surface region and significantly higher than in the adjacent flat area. From these maps we obtained a PL intensity enhancement factor by averaging the PL intensity across the patterned area and dividing it by the average PL intensity from a flat sample region. The enhancement is larger for the transverse magnetic (TM) polarized emission normal to the grooves (~15) than for the TE polarized emission (~5). This is expected as the leaky waves excited by the emitting molecules are longitudinal waves with an electric field in the propagation direction (like SPPs on a smooth metal surface). This field direction is maintained upon decoupling by the HIM. The measured $I_{PL}$ enhancement depends on the groove depth for both TM and TE polarization (Fig. 3e). The ratio between TM and TE emission varies from 1 to 2.5 depending on the groove depth. It is clear that an anisotropic HIM can not only enhance emission, but also control the polarization state of the emitted light directly at the source (without the need for external components like polarizers and wave plates).

To quantitatively understand the origin of this substantial PL enhancement we consider several possible contributions. Under pulsed excitation the measured PL intensity $I_{PL}$ of an ensemble of molecules is proportional to the number of excited molecules per unit area $N$ immediately after the pulse, the collection efficiency of the optics $\eta_{col}$, and the external quantum efficiency $\eta_{ext}$ of the sample:

$$I_{PL} \sim N \times \eta_{col} \times \eta_{ext} \qquad (\text{Eq. 1})$$

We define the external quantum efficiency as:

$$\eta_{ext} = \frac{\Gamma_r}{\Gamma_r + \Gamma_{nr,m} + \Gamma_{nr}^0} = \Gamma_r \times \tau \quad , \qquad (\text{Eq. 2})$$

where $\Gamma_r$ is the radiative decay rate for emission into the far-field either directly or after interacting with the metamirror, $\Gamma_{nr}^0$ is the intrinsic non-radiative decay rate of the molecules, and $\Gamma_{nr,m}$ is the non-radiative decay rate quantifying the dissipation of energy into the metal film. Here, we treat this rate as a non-radiative rate as none of this emission is observed in the far-field (Fig. 3d). The external quantum efficiency can also be written in terms of the measured decay lifetime $\tau = (\Gamma_r + \Gamma_{nr,m} + \Gamma_{nr}^0)^{-1}$ .

In light of Eqs. (1) and (2), an enhancement of emitted intensity can be attributed to an increase in the number of excited dye molecules per collection area, the collection efficiency, and the external quantum efficiency due to changes in radiative and total decay rates. In the rest of this report, we look at the impact of each of these terms.

To compare the excitation of molecules on flat and patterned surfaces, we carried out full-field simulations with different groove depths (Fig. 3f). The absorption of laser light in the emissive layer as a function of groove depth reaches a maximum excitation enhancement factor of ~2.5. Our measurements are in the low-pump power regime (far from saturation). In this regime, the number of excited molecules is linearly proportional to the absorption enhancement.

Regarding the collection efficiency, we can assume that $\eta_{col}$ is essentially the same for patterned and unpatterned areas for two reasons. First, we use a high numerical aperture objective (NA = 0.95). Second, our simulations indicate that the angular radiation pattern of a dipole above the patterned high-impedance metasurface is virtually unchanged compared to a smooth metal surface (see Figs. 1c and S3).

To understand the changes in external quantum efficiency, which depends on the radiative and total decay rates, we measure PL decay traces (Fig. 3g). The traces for the HIM and for a flat silver mirror both show a largely single-exponential decay, indicating that the PL signal predominantly originates from molecules with a single, fast decay rate. Despite the large observed PL enhancement for some groove depths, the decay rate is found to vary by less than 10% for different groove depths. We attribute the small change in lifetime of the emitters to a relatively high intrinsic non-radiative decay rate of R6G molecules when embedded in our chosen polymer matrix. Indeed, the same molecules on a glass substrate have a fast rate of 2.5 ns$^{-1}$, close to 2.9 ns$^{-1}$ measured on an unpatterned silver mirror.

It is thus clear that the small changes in the excitation efficiency (~2.5) and insignificant changes in the collection efficiency and PL decay rate cannot explain the large changes in the PL intensity (~15). These changes can therefore largely be attributed to an enhancement in the radiative rate, *i.e.* a redirection of the emission into the far-field instead of into the metal as expected for a high-impedance surface that suppresses surface plasmons. It should be noted that unlike photonic bandgap structures in dielectrics where the local density of optical states goes to zero inside the bandgap, HIMs trade bound surface modes with leaky modes. As a result, an emitter can couple its emission to these leaky modes, which eventually contribute to the farfield radiation. The enhancement in the radiative rate can be assessed by dividing the measured PL intensity enhancement by the simulated absorption enhancement (Fig. 3h): a maximum enhancement of the radiative decay rate of ~ 5 can be reached with a groove depth of 100 nm.

A periodic array of subwavelength grooves can thus increase the PL intensity by increasing the external quantum efficiency of the emissive layer, while controlling the polarization state of the emission. However, a polarization-independent enhancement in the PL intensity is also possible. To this end, we fabricated high-impedance surfaces based on dimples arranged on a square lattice (Fig. 4). Band structure calculations of such a polarization independent high-impedance surface predict a clear band gap that spans a broad wavelength range 485 - 600 nm (Supplementary Fig. S6). We used the same fabrication process as for the linear grooves to fill the dimples with $SiO_2$ and deposit an emissive layer on top of the high-impedance surface. Again, PL maps demonstrate PL enhancement, but in this case the enhancement is independent of the detection polarization. The enhancement increased with dimple depth and reaches a maximum value of ~12 for a groove depth of 180 nm (Fig. 4d), similar to the maximum TM polarization value for grooved samples but now obtained for both polarizations. Similar to the linear groove arrays, the lifetime measurements do not show any significant decay rate differences for the molecules on the dimple arrays compared to those on a flat sample region. We conclude that similar physics underlies the observed PL enhancement for both dimple and groove arrays.

In conclusion, motivated and inspired by the notion of high-impedance ground planes in the radio frequency regime, we have developed an optical version of this concept for enhancing optical emission, demonstrating the realization of high electrical conductivity metallic electrodes that also have high reflectivity at visible frequencies and that are free of SPP loss associated with traditional metallic electrodes. These high-impedance metasurfaces can enhance the light extraction efficiency of quantum emitters and reduce heat generation in metallic electrodes by introducing a wide SPP band gap. Within the gap no guided surface waves exist and the emission is rapidly decoupled to free-space. As the relative fractions of the power coupled to SPPs versus free space on a smooth metal just depends on the distance of the emitter to the metal and not on the internal quantum efficiency, this approach is effective for both low and high internal-quantum-efficiency emitters. The broadband nature of the SPP band gap observed for these metasurfaces makes them promising candidates to improve the performance of solid-state light-emitting devices for illumination and display applications. Moreover, these metasurfaces provide an excellent platform for engineering and manipulating SPPs along two-dimensional structures (flat photonics), allowing additional functionality in light-matter interaction.

**Methods:**

**Device fabrication.** High-impedance metasurfaces were fabricated in a multistep process. Silicon wafers with 300-nm-thick thermally grown oxide were used as a planar and smooth substrate. They were cleaned in several steps. First, they were sonicated with Alconox precision cleaner solution, then rinsed with water three times. After this step, they were sonicated for 5 minutes in Acetone, Methanol, and Isopropanol, consecutively, and finally dried with nitrogen gas. A smooth, optically thick (300 nm) silver film was then deposited onto the substrate using a 2 nm germanium nucleation layer. The metal was nanopatterned by FIB milling with a FEI Helios 600i dual FIB/SEM tool. A thick layer of silicon oxide was deposited on the patterned area with electron beam assisted deposition for the first 50 nm and ion beam assisted deposition for the next 250 nm. Then the deposited $SiO_2$ layer was milled down to the silver surface with FIB until a 5~10 nm overcoat of $SiO_2$ was left on the high-impedance area and the surrounding flat area.

To deposit a thin light-emitting layer, Rhodamine 6G powder was dissolved in a Polymethyl methacrylate (PMMA) solution. The PMMA was diluted with Anisole to reduce the achievable thickness of the emissive layer. The solution was then put in a vortex mixer for 2 minutes. The concentration of R6G in the PMMA solution is approximately $3\times10^{-4}$ M. The thickness of the R6G film is around 15 nm, as measured by SEM imaging the cross section of our samples.

**Optical measurements.** PL maps were acquired with a Witec alpha 300 R confocal microscope with a two-axis piezo stage for sample scanning. A PicoQuant LDH-P-C-485 pulsed laser at wavelength of 485 nm is used for excitation of R6G molecules. The PL signal is detected with an avalanche photodiode (MPD APD) for both lifetime measurement and PL intensity measurement. The spatial resolution of the imaging system in our experiments is about 500 nm. The measured lifetime traces were fitted with a bi-exponential function of the form $I(t) = ae^{-r_1 t} + (1-a)e^{-r_2 t}$ where $a$ is the amplitude of the fast contribution and $r_1$ and $r_2$ are the decay rates of the two contributions. The value of $a$ is larger than 0.99 for all five groove depths (Fig. S5)

**Simulations.** All simulations were carried out using the Finite-Difference Time-Domain method (Lumerical FDTD Solutions).

**Acknowledgements**

This work was supported by a Multi University Research Initiative (MURI FA9550-12-1-0488) from the AFOSR and a gift from Konica Minolta Laboratory USA. A.G.C. also acknowledges the support of a Marie Curie International Outgoing Fellowship.

**Figures**

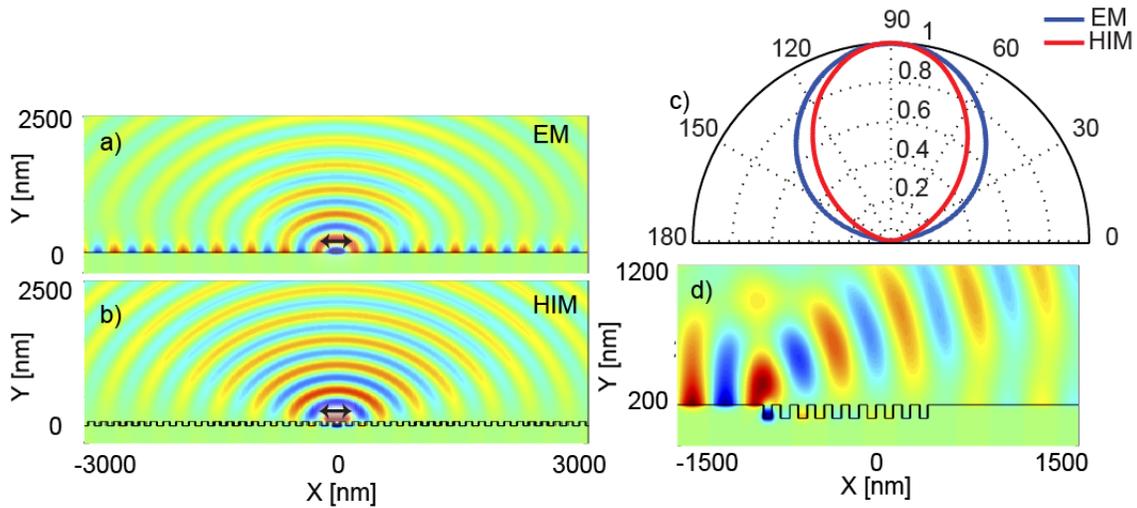

**Fig. 1. Absence of propagating surface plasmon polaritons (SPPs) on a high-impedance metasurface (HIM) metal electrode. a)** Simulation of the magnetic field profile for an electric dipole emitting at wavelength of 600 nm and positioned 10 nm above a smooth silver film serving as an electrical mirror (EM). The orientation of the dipole is parallel to the silver surface. Part of the emission is coupled to surface plasmon polaritons that along the interface which are finally dissipated in the metal. **b)** Magnetic field profile for an electric dipole above a HIM patterned with 100-nm-deep grooves spaced at a subwavelength periodicity of 150 nm and for a metal filling fraction of 50 %. Same dipole orientation as in (a) No emission into SPPs is observed and the radiation into free space is enhanced. **c)** Broad angular emission distribution for a horizontal dipole above a smooth EM and a patterned HIM. (d) Decoupling of SPPs incident on a patterned HIM area consisting of a subwavelength array of grooves with same dimensions as in (b) Excitation wavelength for SPPs is 600nm.

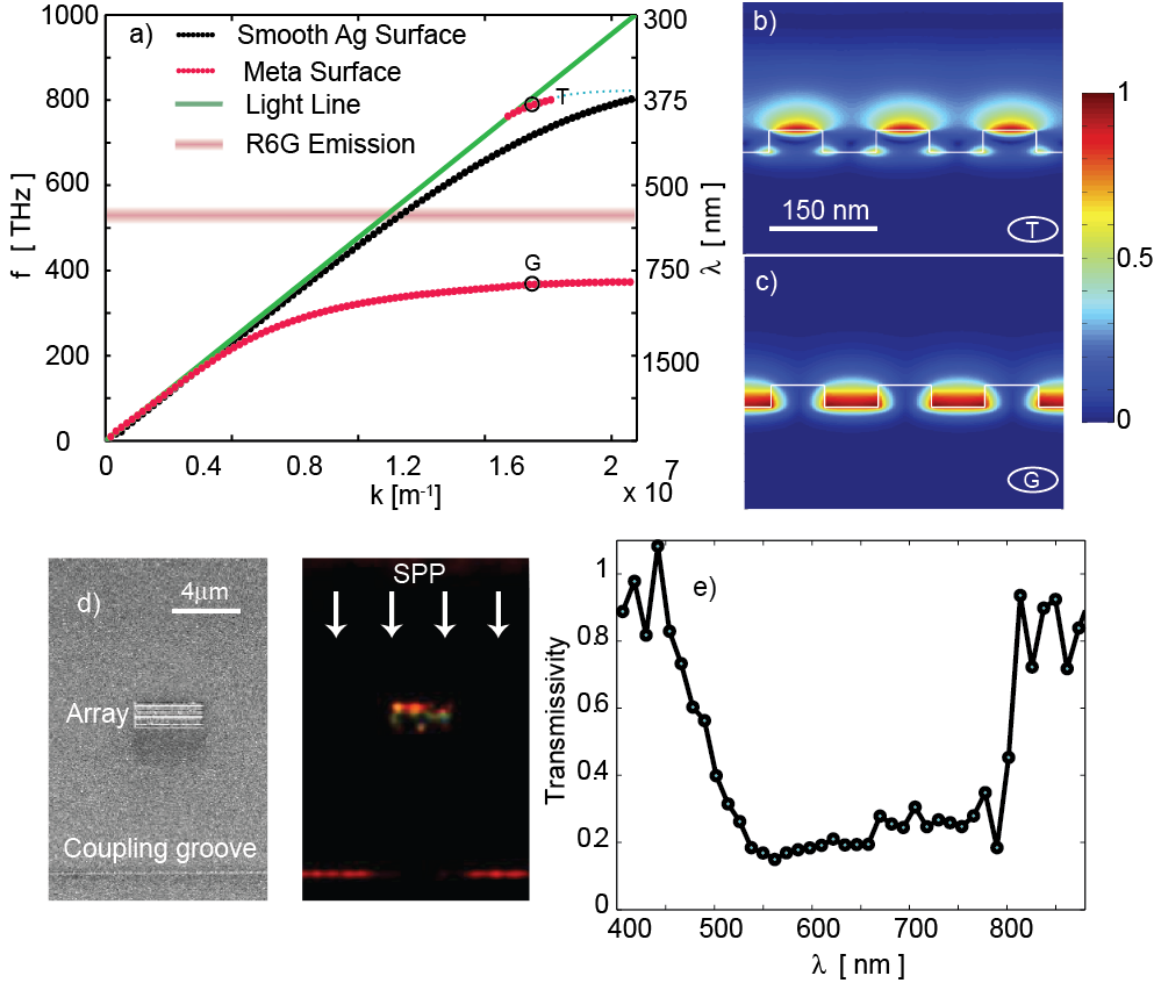

**Fig. 2. Surface plasmon band gap for a high-impedance metasurface consisting of a metal patterned with a subwavelength groove-array. a)** Dispersion relation for SPPs propagating along a smooth silver/air interface (black), a high-impedance metasurface (red) consisting of 100-nm-deep and 75-nm-wide grooves space at period of 150 nm, and the light line in air (green). The spectral emission band of R6G molecules is also shown in the figure demonstrating that dye's emission falls into the bandgap of the high impedance surface. **b-c)** Spatial distribution of magnetic field magnitude $|H_z|$ for the high- and low-frequency modes of the high-impedance metasurface at the points labeled T (for top) and G (for gap) in the dispersion relation. **d)** SEM (left) and optical (right) images. The groove array is illuminated with a planar SPP wave from the top (arrows). The groove array decouples the SPPs to the far field and casts a shadow on the outcoupling groove at the bottom of the image **e)** Measured transmission spectrum for the SPPs across the groove array, defined as the ratio of light intensity scattered at the exit slit with and without the array.

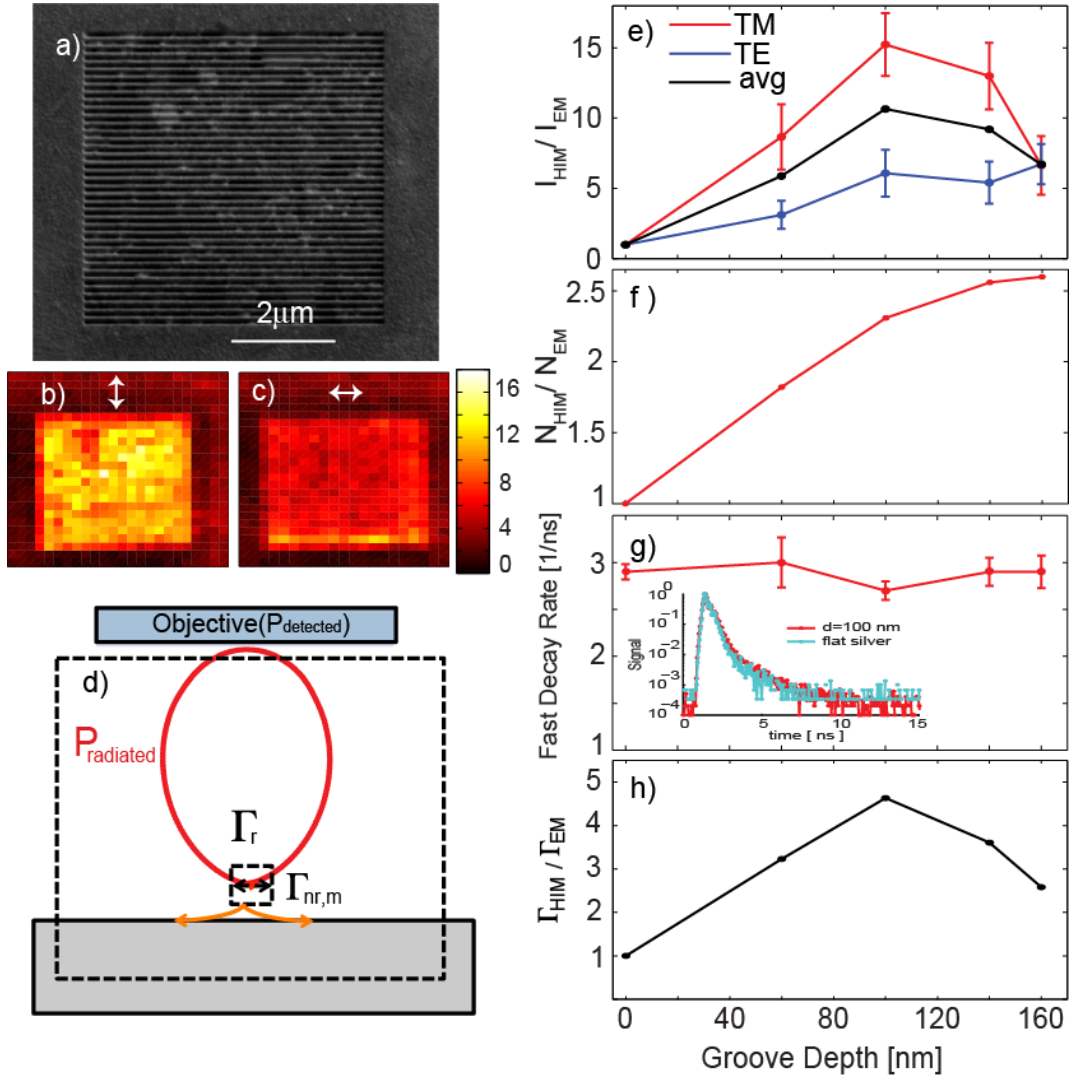

**Fig. 3. Photoluminescence intensity enhancement on a high-impedance metasurface.**
**a)** SEM image of 100-nm-deep groove-array patterned into a smooth Ag film. **b-c)** Photoluminescence maps of a grooved surface covered with a R6G dye layer. The excitation laser is polarized along the grooves. The detection polarization is perpendicular or parallel to the grooves, respectively. **d)** Schematic of different quantities used in our analysis of the PL decay. **e)** PL intensity enhancement as a function of the groove depth for TM and TE polarized collection, as well as the average. **f)** Simulated enhancement of the density of excited molecules in the emissive layer as a function of groove depth. **g)** Dependence of the dominant fast-decay component of the PL lifetime for molecules on top of a smooth silver film and on HIMs with different groove depth. The inset shows examples of measured lifetime traces for the case of groove depth of 100 nm and flat mirror. **h)** Calculated radiative decay rate enhancement. Error bars indicate the 95% confidence interval of the fit.

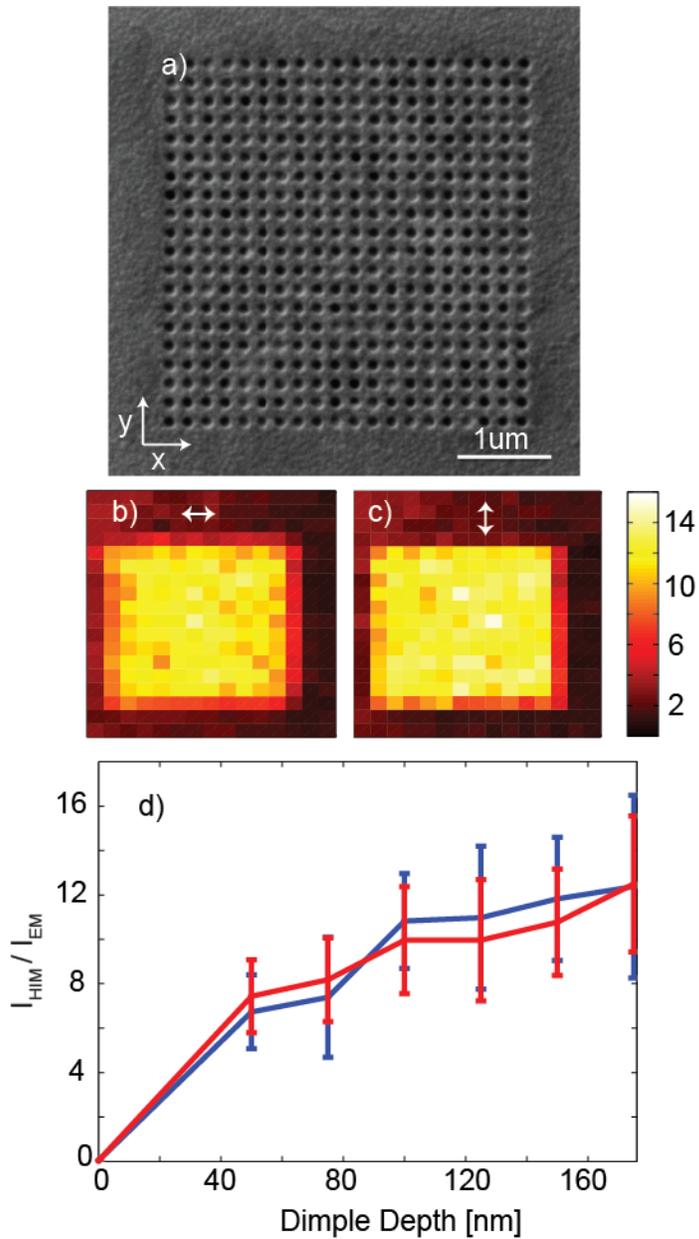

**Fig. 4. Polarization-independent photoluminescence enhancement achieved with a high-impedance metasurface featuring an array of nanodimples a)** SEM image of a subwavelength dimple array patterned into a silver substrate. **b-c)** PL maps of molecules placed on a HIM with nanodimples and for two orthogonal polarization directions for the optical collection and a fixed excitation polarization along the *x*-axis. d) Measured PL intensity enhancement for polarizations along the *x*-axis (blue) and *y*-axis (red) as a function of dimple depth.

# Supplementary information materials

## Optical emission near a high-impedance mirror

Majid Esfandyarpour, Alberto G. Curto, Pieter G. Kik, Nader Engheta,

and Mark L. Brongersma

## Table of Contents

### Sections

### Supplementary Figures

**Section 1. Emission properties of a dipole near a smooth and a patterned surface**

In this section we use simulations to compare the coupling of an electric dipole emitter to surface plasmon polariton (SPP) modes, when it is placed above a smooth silver (Ag) film and when it is placed above a high-impedance metasurface.

Fig S1a shows the real part of the magnetic field profile ($H_z$) for an electric dipole placed 10 nm above the surface of a flat Ag film with an orientation perpendicular to the metal surface. The field profile clearly shows that the emitted photons are efficiently coupled to SPP modes that travel along the glass/Ag interface. The emission wavelength is 600 nm. If we replace the flat silver mirror by a grooved metasurface with a periodicity of 150 nm, a groove depth of 100 nm and a filling fraction of 50%, we do not observe propagating SPPs along the surface. Instead, the dipole is emitting more efficiently to the far field, as shown in Fig. S1b. We also simulated the emission of an electric dipole with orientation parallel to the metal surface for the cases of a smooth and corrugated metal surface. The real part of the magnetic field profiles for these cases are shown in Fig S1c and Fig S1d respectively. For this dipole, a qualitative similar behavior is observed.

A more quantitative way of analyzing the effect of a high-impedance metasurface on the far-field emission of a dipole is by quantifying the fraction of the radiated power that is absorbed by the metal. To do so, we simulate the emission of an electric dipole emitting at wavelength of 560 nm placed in glass (refractive index of 1.5) above a high-impedance metasurface consisting of silver patterned with a subwavelength array of grooves with periodicity of 150 nm , width of 75 nm and depth of 100 nm. We compare it to a flat silver mirror. The simulations are done by commercial Finite-Difference Time-Domain package (Lumerical FDTD Solutions). The distance between the dipole and mirror is changed from 10 to 160 nm. There are two power monitors below and above the dipole measuring the power absorbed in the metal (undesired) and the total radiated to the far field (desired) respectively. The simulation size is chosen to be relatively large (40 μm x 40 μm) to ensure complete decoupling/dissipation of SPP into the metal within the simulation volume. We define the loss ratio as the power going through the monitor below the dipole to the total radiated power by the dipole. Figure S2a shows the loss ratio for an electric dipole oriented parallel to the surface of a smooth Ag surface as a function of dipole distance from the mirror surface. The loss ratio is substantially lower for the case of the high-impedance metasurface when the dipole distance is less than 60 nm. The peak in the loss ratio occurs at a distance of 80 nm for the metasurface due to the fact that the dipole is now placed close to the antinode of a standing wave profile created by the reflection of a normally-incident plane wave from the surface. Figure S2b shows the loss ratio as a function of the dipole distance for a dipole orientation perpendicular to the metal surface. For this orientation of dipole coupling of radiated photons to SPP modes is more efficient. We can understand this intuitively by looking at the angular emission as shown in Fig. S3b, which is more directed along the surface. As a result, the loss ratio for this dipole orientation is above 80% for dipole distances smaller than 120 nm when placed above a smooth Ag film. If we replace the smooth Ag mirror by a high-impedance metasurface the loss ratio is dropped below 10% for all dipole distances from 10 to 160 nm.

**Section 2: SPP transmission and reflection across a finite number of grooves**

In this section we study the transmission of SPP waves across 10 deep subwavelength grooves with 150 nm periodicity and 50% filing factor. A SPP wave source at a wavelength of 560 nm is placed on the left side of the grooves. SPP waves travel along the silver/air interface and are decoupled to far-field radiation by means of grooves. Transmission and reflection coefficients of SPP waves are calculated as a function of grooves depth and shown in Fig. S5a-b, respectively. The broadband nature of SPP decoupling by grooves can be seen from this figure. For example, at a groove depth of 80 nm the transmission is below 20% for wavelengths between 400 and 800 nm and the reflection amplitude is also less than 10%.

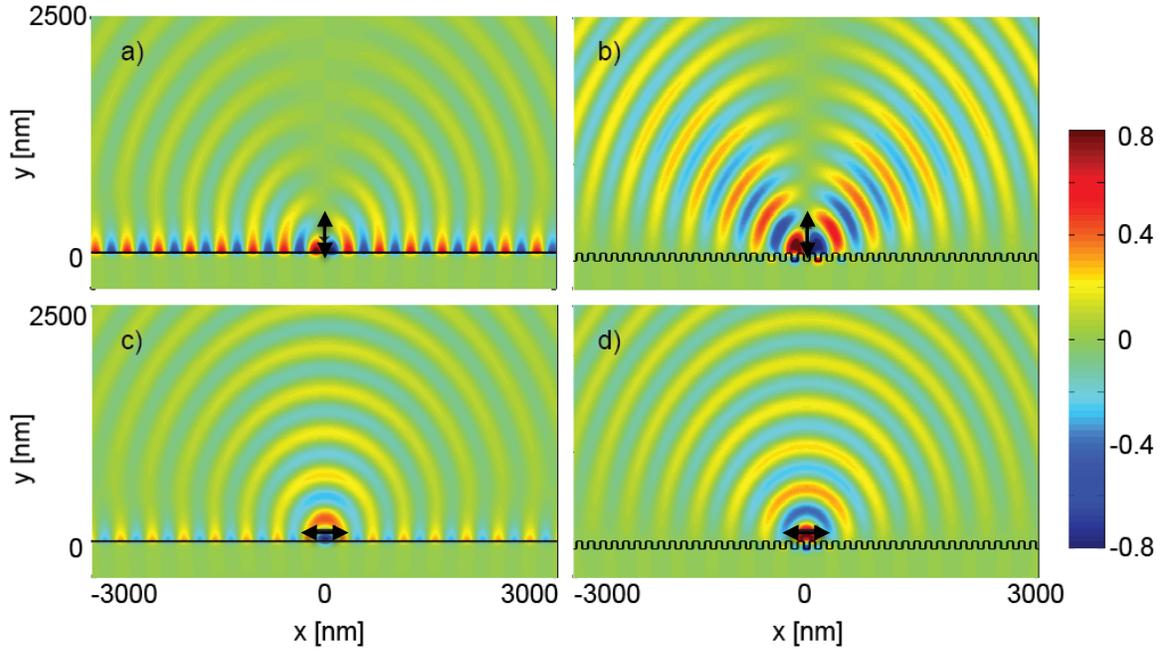

**Fig. S1. Optical simulation of electric dipole emission near smooth and patterned metal surfaces.** a) Normalized real part of magnetic field profile ($H_z$) of an electric dipole emission oriented perpendicular to the surface of a flat Ag mirror and b) metamirror c) Dipole oriented parallel to the surface of a flat Ag mirror, and d) metamirror.

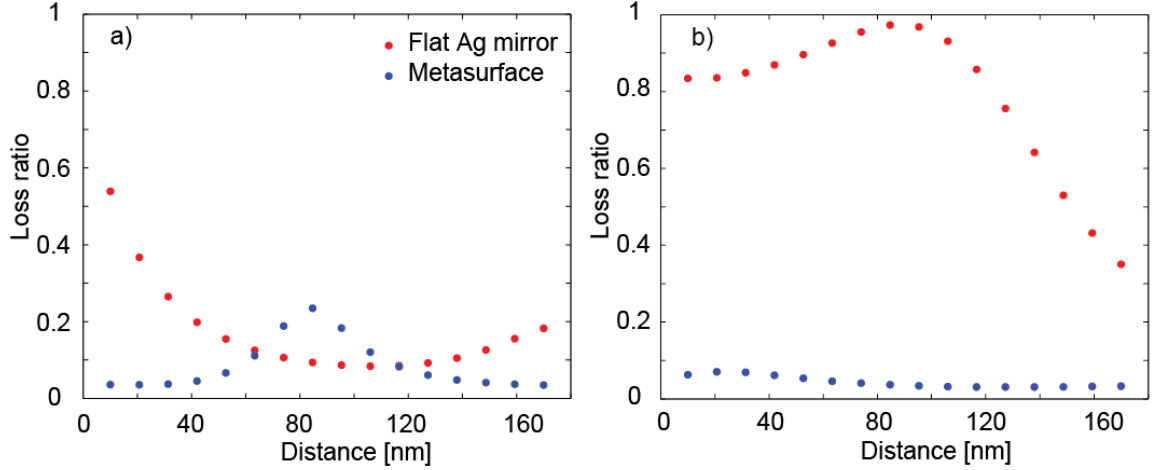

**Fig. S2. Dependence of the loss ratio for electric dipoles above a flat Ag film and a high impedance metasurface.** a) Calculation for an electric dipole emitter oriented parallel to the surface and b) Same calculation for a dipole oriented perpendicular to the surface. High impedance metasurfaces consist of grooves with depth of 100 nm and a filling fraction of 50%. The periodicity is 150nm.

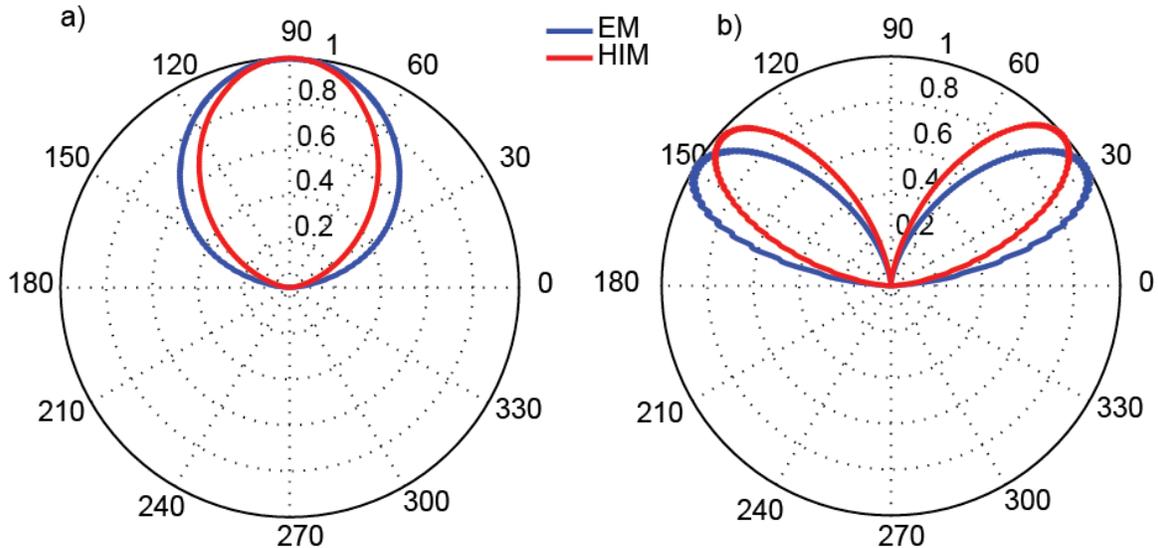

**Fig. S3. Far-field angular radiation pattern for an electric dipole above a smooth electric mirror (EM) and a high-impedance metasurface (HIM).** a) Dipole oriented parallel to the surface of mirror. b) Dipole oriented perpendicular to the surface of the mirror. The dipole distance from the surface is 10nm.

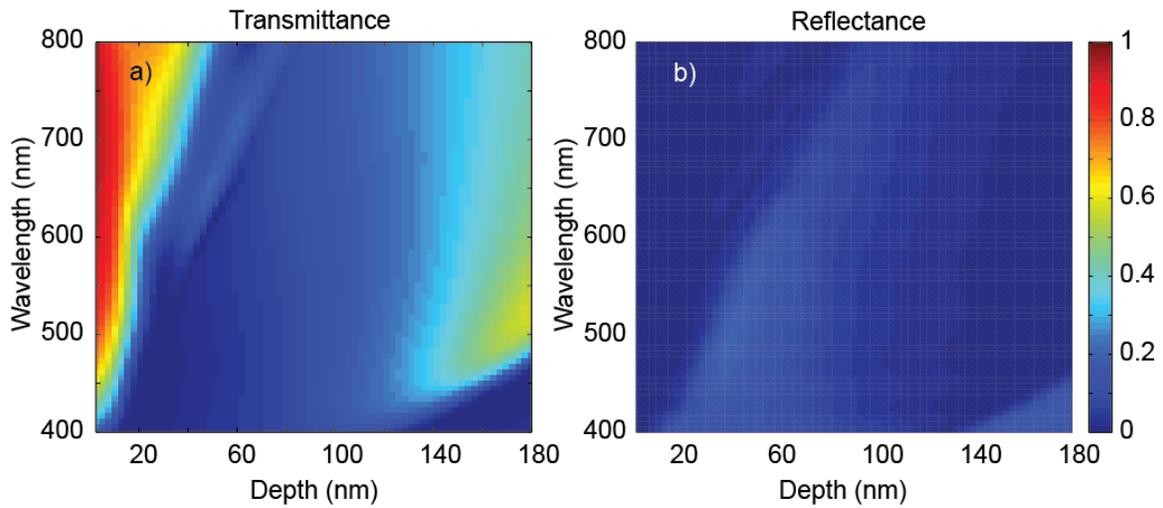

**Fig. S4. Optical simulation of the surface plasmon transmission and reflection across a finite number of subwavelength grooves.** a) Maps of the calculated transmittance and b) reflectance of SPPs across a series of 10 grooves with the same dimensions as the grooves in Figure 2 of the main text (periodicity of 150 nm, and groove width of 75 nm) The vertical axis shows the free space wavelength and the horizontal axis provide the roove depth.

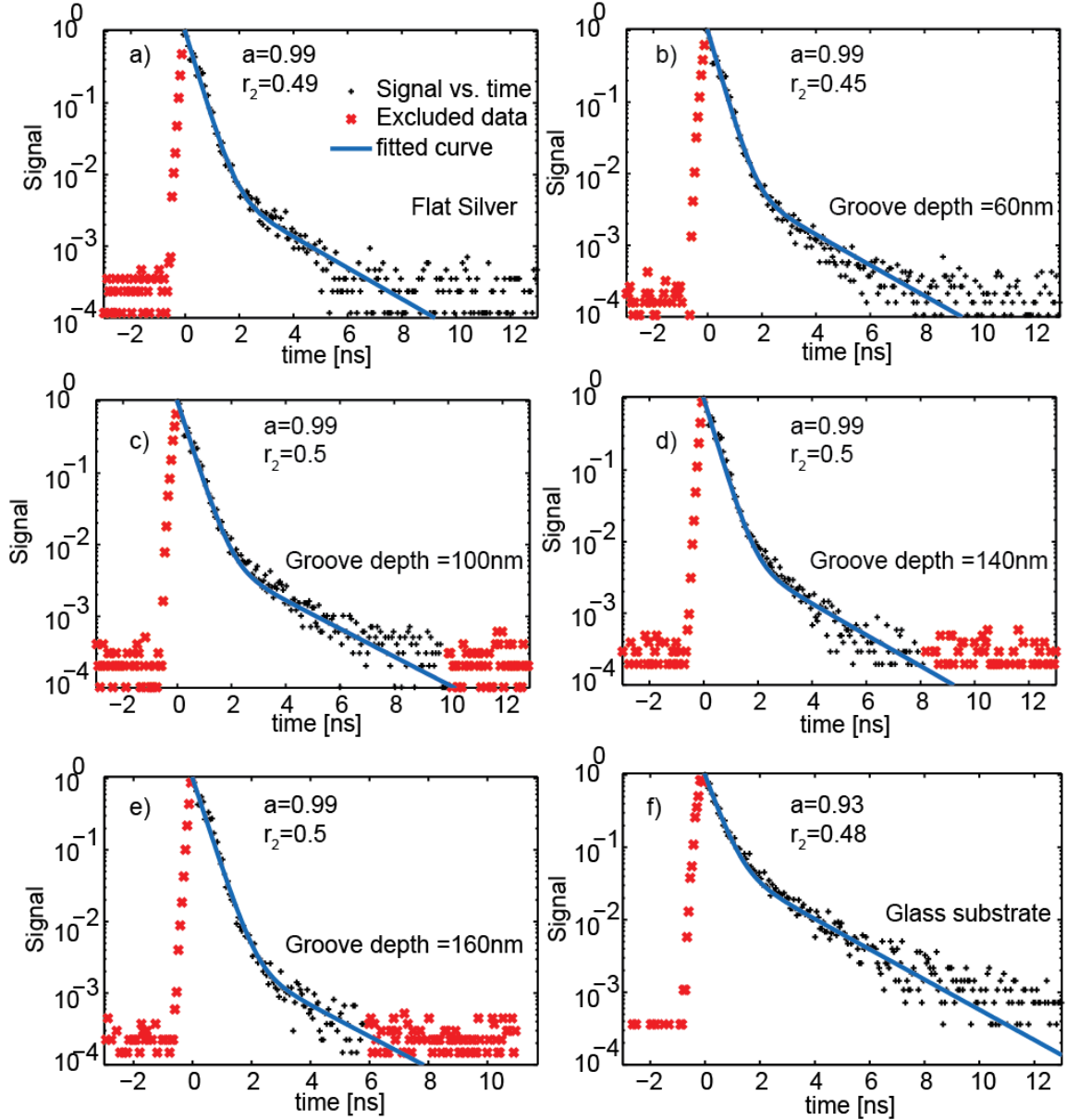

**Fig. S5. Time-resolved photoluminescence measurement of R6G dye molecules above different metallic surfaces.** Time-resolved decay of 15-nm-thick R6G dye molecule with 5~10 nm spacer layer between dye layer and surface for **a)** flat silver surface **b)** metasurface with groove depths of 60 nm **c)** depth of 100 nm **d)** depth of 140 nm **e)** depth of 160 nm. **f)** Time-resolved measurement of R6G dye on a glass substrate. Each time trace is fitted with a double exponential function of the form $I(t) = ae^{-r_1 t} + (1-a)e^{-r_2 t}$, where $I$ is the signal, $r_1$ and $r_2$ are the fast and slow rates, respectively, and $a$ is the amplitude of the fast decay.

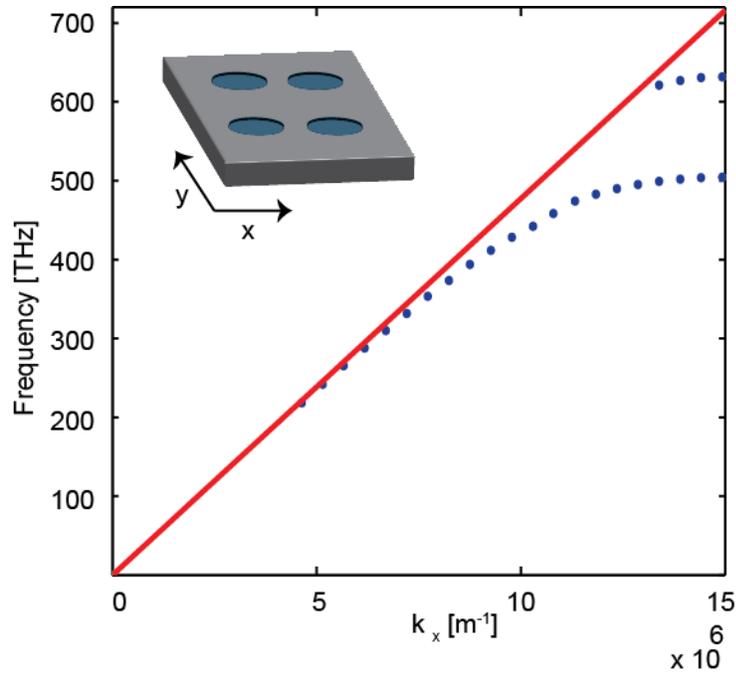

**Fig. S6. Band structure calculation for a periodic dimple array.** Band structure of a periodic dimple array with periodicity of 210 nm, a SiO2-filled dimple depth of 120 nm and a radius of 65 nm, along with the light line showing a band gap for a broad wavelength range from 485 to 600 nm.